\documentclass[a4paper,twocolumn,aps,eqsecnum]{revtex4}

\usepackage{bm}% bold math
\usepackage{amsmath}
\usepackage{amssymb}
\usepackage{amsfonts}
\usepackage[dvips]{color}
\newcommand{\half}{\frac{1}{2}}

\newcommand{\der}{\partial}

\newcommand{\bfk}{\bm{k}}

\newcommand{\bfp}{\bm{p}}

\begin{document}

% Use the \preprint command to place your local institutional report
% number in the upper righthand corner of the title page in preprint mode.
% Multiple \preprint commands are allowed.
% Use the 'preprintnumbers' class option to override journal defaults
% to display numbers if necessary
\preprint{KYUSHU-HET-88}

%Title of paper
\title{Wilsonian RG and Redundant Operators \\
in Nonrelativistic Effective Field Theory}

% repeat the \author .. \affiliation  etc. as needed
% \email, \thanks, \homepage, \altaffiliation all apply to the current
% author. Explanatory text should go in the []'s, actual e-mail
% address or url should go in the {}'s for \email and \homepage.
% Please use the appropriate macro foreach each type of information

% \affiliation command applies to all authors since the last
% \affiliation command. The \affiliation command should follow the
% other information
% \affiliation can be followed by \email, \homepage, \thanks as well.
\author{Koji Harada}
\email{koji1scp@mbox.nc.kyushu-u.ac.jp}
\author{Kenzo Inoue}
\email{inou1scp@mbox.nc.kyushu-u.ac.jp}
\author{Hirofumi Kubo}
\email{kubo@higgs.phys.kyushu-u.ac.jp}
%\email[]{koji-kun@quark.phys.kyushu-u.ac.jp}
%\homepage[URL: ]{http://higgs.phys.kyushu-u.ac.jp/koji}
%\thanks{}
%\altaffiliation{}
\affiliation{Department of Physics, Kyushu University\\
Fukuoka 812-8581 Japan}

%Collaboration name if desired (requires use of superscriptaddress
%option in \documentclass). \noaffiliation is required (may also be
%used with the \author command).
%\collaboration can be followed by \email, \homepage, \thanks as well.
%\collaboration{}
%\noaffiliation
\date{\today}

\begin{abstract}
 In a Wilsonian renormalization group (RG) analysis, redundant
 operators, which may be eliminated by using field redefinitions, emerge
 naturally. It is therefore important to include them. We consider a
 nonrelativistic effective theory (the so-called ``pionless'' Nuclear
 Effective Field Theory) as a concrete example and show that the
 off-shell amplitudes cannot be renormalized if the redundant operators
 are not included. The relation between the theories with and without
 such redundant operators is established in the low-energy expansion. We
 perform a Wilsonian RG analysis for the \textit{off-shell} scattering
 amplitude in the theory with the redundant operator.

%We emphasize that the Jacobian for a nonlinear field
% redefinition is in general nontrivial, and should be taken into account
% in a Wilsonian RG analysis.
\end{abstract}

% insert suggested PACS numbers in braces on next line
% insert suggested keywords - APS authors don't need to do this
%\keywords{}

%\maketitle must follow title, authors, abstract, \pacs, and \keywords
\maketitle

% body of paper here - Use proper section commands
% References should be done using the \cite, \ref, and \label commands

%\tableofcontents

\section{Introduction}

Effective field theories\cite{Weinberg:1978kz, Lepage:1989hf,
Polchinski:1992ed, Georgi:1994qn, Kaplan:1995uv, Manohar:1996cq} are
useful in exploring the low-energy physics with only relevant degrees of
freedom and operators. Chiral Perturbation Theory
($\chi$PT)\cite{Gasser:1983yg} is a prominent example, in which only
meson degrees of freedom are treated. The idea of $\chi$PT has been
applied to the sectors with baryons. The Nuclear Effective Field Theory
(EFT)\cite{Weinberg:1990rz, Weinberg:1991um} is that with two baryons
and more. The Nuclear EFT is a promising ``model-independent'' approach
to nuclear physics, based on general principles of quantum field theory.
Because of its inherent nonperturbative nature, however, we are still
unable to understand the basic structure of the theory, though some
preliminary successes have been reported. See Refs.\cite{Beane:2000fx,
Bedaque:2002mn} for reviews.

One of the most important issues is how to determine the power counting.
There are mainly two distinct power counting schemes: the one due to
Weinberg\cite{Weinberg:1990rz, Weinberg:1991um} is based on the naive
dimensional analysis for the construction of the ``effective potential''
which is used in the Lippmann-Schwinger equation, and the other proposed
by Kaplan, Savage, and Wise\cite{Kaplan:1998tg,Kaplan:1998we} takes into
account that the scattering length is unnaturally long and is
implemented by the so-called \textit{power divergence subtraction}. It
has been shown that Weinberg's power counting is
inconsistent\cite{Kaplan:1996xu} while the KSW power counting fails to
converge in certain channels\cite{Fleming:1999ee}. The present status
may be summarized as that Nuclear EFT works ``pretty well if one follows a
patchwork of power counting rules\cite{Kaplan:2005es}.'' Clearly, a
\textit{principle} of systematizing the power counting, which tells the
correct one \textit{before} doing numerical calculations, is
needed. (See Ref.\cite{Beane:2001bc} for a possible solution.)

We think that such a principle may be provided by a Wilsonian
renormalization group (RG) approach\cite{Wilson:1973jj,Wilson:1974mb}. In
a Wilsonian approach, one does not need to prescribe a particular power
counting. We hope that the RG flow itself determines how to treat the
operators.

A Wilsonian RG approach has been performed by Birse and
collaborators\cite{Birse:1998dk, Barford:2002je}.  They consider the RG
equation for a \textit{energy-} (as well as momentum-)
\textit{dependent} potential.  (See Ref.\cite{Bogner:2003wn} for a
similar but distinct approach and Ref.\cite{Nakamura:2005uw} for the
comparison.) Note that the notion of Wilsonian RG should be independent
of the on-shell nature.  Note also that the potential approach has the
so-called off-shell ambiguities, which may be a problem when one
considers the three-nucleon systems. A completely field theoretical
treatment is desired. We have been developing it, and will be reported
elsewhere\cite{HK}. In this paper, we will give some preparatory remarks
that are indispensable for understanding the development.

In a completely field theoretical treatment, a Wilsonian RG transformation
generates all kinds of operators consistent with the symmetry. Among
them, there are operators which are proportional to the (tree-level)
equations of motion. These operators, which we call ``redundant,'' are
usually eliminated by a field redefinition. In order for the RG
equations not to contain the couplings for such redundant operators, 
the field redefinition should follow the RG transformation.

It is well-known that the physical S-matrix is independent of the choice
of the field variables\cite{Kamefuchi:1961sb}. One might therefore think
that the field redefinition mentioned above is a trivial procedure. In
this paper, we illustrate that it is far from trivial. We consider a
simple nonrelativistic system (known as ``pionless'' EFT) as a concrete
example and show how the redundant operator is eliminated without
affecting physics.  (Note that we do not claim that pionless EFT is
particularly interesting. As easily seen, our results are very general
and the simplest example illustrates the essential points.)  We show
that the coupling constants are transformed nontrivially. With this
effect taken into account, we obtain the RG equations for the reduced
set of coupling constants.

In Sec.~\ref{sec:indep}, we review the standard argument for the
independence of the physical S-matrix on the choice of the field
variables, and emphasize that there are contributions from the measure.
In Sec.~\ref{sec:NEFT} we give an explicit calculation of the two
nucleon scattering amplitude in the pionless Nuclear EFT, and show that
the off-shell amplitude for the theory without the redundant operator
cannot be renormalized. We further show that it is possible to eliminate
redundant operators without modifying physics, but the coefficients of
the other operators must be modified in a nontrivial way. In
Sec.~\ref{sec:RG} we derive the RG equations and renormalize the
\textit{off-shell} scattering amplitude for the theory with the
redundant operator.  In Sec.~\ref{sec:conclusion} we summarize the
results.

\section{Independence of the physical S-matrix on the choice of the field variables}
\label{sec:indep}

It is well known that the physical S-matrix elements are independent of
the choice of the field variables\cite{Kamefuchi:1961sb}.  A standard
argument goes as
follows\cite{Kallosh:1972ap,Bando:1987br,Manohar:1998re}. Consider a
scalar field theory for simplicity. We consider the following point
transformation,
\begin{equation}
 \phi \rightarrow \Phi\equiv f(\phi),
\end{equation}
and the two $n$-point functions,
\begin{eqnarray}
 G_{(n\phi)}(x_1,\cdots, x_n)&=&\langle T \phi(x_1)\cdots
  \phi(x_n)\rangle, \\
 G_{(n\Phi)}(x_1, \cdots, x_n)&=&\langle T \Phi(x_1)\cdots
  \Phi(x_n)\rangle \nonumber \\
 &=&\langle T f(\phi)(x_1)\cdots f(\phi)(x_n)\rangle.
\end{eqnarray}
According to the LSZ formalism, the S-matrix elements in the
$\Phi$-theory are obtained by multiplying the Klein-Gordon operator
$\Box+m^2$ (where $m$ is the mass of the particle) to each leg of the
connected part of $G_{n\Phi}$ and considering the on-shell limit
$p^2\rightarrow m^2$. Because only the single particle propagation has
the pole, this procedure picks up the $\Phi$-$\phi$ transition part of
all the possible diagrams, giving rise to the same S-matrix elements. (A
possible wave function renormalization factor is canceled anyway in
obtaining the S-matrix elements.)

The above argument itself is of course valid but it might cause some
confusion in practice. The point is best illustrated in the path
integral formulation. The $n$-point function $G_{(n\phi)}$ may be given
by
\begin{equation}
 G_{(n\phi)}(x_1,\cdots, x_n)=\int {\cal D}\phi \phi(x_1)\cdots
  \phi(x_n)
  e^{i\int d^4x {\cal L}(\phi)},
\end{equation}
while $G_{(n\Phi)}$ by
\begin{eqnarray}
 &&G_{(n\Phi)}(x_1,\cdots,x_n) \nonumber \\
  &=&\int {\cal D}\phi f(\phi)(x_1)\cdots
  f(\phi)(x_n)e^{i\int d^4x {\cal L}(\phi)},
\end{eqnarray}
not by 
\begin{equation}
 \int {\cal D}\Phi \Phi(x_1)\cdots \Phi(x_n)
  e^{i\int d^4x \tilde{\cal L}(\Phi)},
\end{equation}
where 
\begin{equation}
 \tilde{\cal L}(f(\phi))={\cal L}(\phi).
\end{equation}
The difference comes from the Jacobian factor, $\det\left|{\cal
D}f(\phi)/{\cal D}\phi\right|^{-1}$, which introduces new interactions
in the $\Phi$-theory.

In a fermionic theory, $\psi$ and $\psi^\dagger$ are essentially a
canonically conjugate pair. A field transformation which mixes $\psi$
and $\psi^\dagger$ is not a point transformation nor (generally) a
canonical transformation, and thus leads to a nontrivial Jacobian factor.
As we will see in the following section, the field redefinition which
eliminates redundant operators is of this kind.

In many cases, the Jacobian may be disregarded\cite{Georgi:1991ch,
Arzt:1993gz}. In perturbation theory it gives higher order
contributions, and when dimensional regularization is employed it does
not contribute at all.  In any case, the Jacobian may be represented as
a series of local interactions\footnote{If the field transformation
contains derivatives, the Jacobian may be represented as a ``ghost''
term which generates nonlocal interactions. We do not consider such a
case in this paper.}, which are already present in EFT, so
that it is absorbed in the definitions of the coupling constants.  The
nonrelativistic theory that we will consider in the next section,
however, allows us to calculate the amplitudes nonperturbatively, and in
order to perform a Wilsonian RG analysis, dimensional regularization
should not be employed. An explicit calculation reveals the nontrivial
character of the contributions from the measure.

\section{Pionless Nuclear EFT as a concrete example}
\label{sec:NEFT}

We consider the following simple nonrelativistic effective field theory,
\begin{eqnarray}
 {\cal L}&=&N^\dagger\left(i\der_t+\frac{\nabla^2}{2M}\right)N 
  -C_0\left(N^TN\right)^\dagger(N^TN)
  \nonumber \\
 &&{}+ C_2
  \left[
   \left(N^TN\right)^\dagger\left(N^T\overleftrightarrow{\nabla}^2N\right)
   + h.c.
  \right] + \cdots
\end{eqnarray}
where $N$ is a ``spinless nucleon'' field\cite{Kaplan:1998we}, 
\begin{equation}
 \overleftrightarrow{\nabla}^2=
  \overrightarrow{\nabla}\cdot\overrightarrow{\nabla}
  -2\overleftarrow{\nabla}\cdot\overrightarrow{\nabla}
  +\overleftarrow{\nabla}\cdot\overleftarrow{\nabla},
\end{equation}
and ellipsis denotes higher order operators. Though the spin is
neglected in this theory, the results are identical to those for the
spin singlet channel. Note that, because of the nonrelativistic nature,
the production of anti-particles is suppressed and the particle number
is conserved. In particular, the six-nucleon operators, such as
$\left(N^\dagger N\right)^3$ do not contribute in the two particle
sector. Note also that the theory is invariant under Galilean boost. We
have to keep this symmetry.

The Lippmann-Schwinger (LS) equation for the two particle scattering
amplitude ${\cal A}(p^0,\bfp_1, \bfp_2)$ in the center-of-mass frame is
given by
\begin{eqnarray}
 &&\!\!\!-i{\cal A}(p^0,\bfp_1, \bfp_2)=-iV(\bfp_1, \bfp_2) \nonumber \\
 &&\!\!\!{}+\!\!\int \!\!\!\frac{d^3k}{(2\pi)^3}
  \left(-iV(\bfk,\bfp_2)\right)\!
  \frac{i}{p^0\!-\!\bfk^2/M\!+\!i\epsilon}
  \left(\!-i {\cal A}(p^0,\bfp_1,\bfk)\right), \nonumber \\
\end{eqnarray}
where $V$ is the vertex in momentum space,
\begin{equation}
 V(\bfp_1,\bfp_2)=C_0+4C_2\left(\bfp_1^2+\bfp_2^2\right)+\cdots,
\end{equation}
and $p^0$ is the center-of-mass energy of the system, $\bfp_1$ and
$\bfp_2$ are the relative momenta in the initial and final states
respectively. We consider the off-shell amplitude for the moment so that
$p^0$, $\bfp_1$, and $\bfp_2$ are unrelated.

By introducing a cutoff $\Lambda$ on the relative momentum $\bfk$ and
expanding the amplitude up to including quadratic terms in each momentum,
\begin{equation}
 {\cal A}(p^0,\bfp_1,\bfp_2)=x(p^0) 
  + y(p^0)(\bfp_1^2+\bfp_2^2)+z(p^0) \bfp_1^2\bfp_2^2,
\end{equation}
we may solve the LS equation in a closed form\cite{Gegelia:1998iu,
Phillips:1997xu}:
\begin{eqnarray}
 x&=&\left(C_0+16C_2^2I_2\right)/D, \\
 y&=&4C_2\left(1-4C_2I_1\right)/D, \\
 z&=&16C_2^2I_0/D,
\end{eqnarray}
with
\begin{equation}
 D=1-C_0I_0-8C_2I_1+16C_2^2I_1^2-16C_2^2I_0I_2,
\end{equation}
where $I_n$ are integrals defined by
\begin{equation}
 I_n=-\frac{M}{2\pi^2}\int_0^\Lambda dk \frac{k^{2n+2}}{k^2+\mu^2},
  \quad \mu=\sqrt{-Mp^0-i\epsilon}.
\end{equation}

At low energies, we can expand the inverse of the \textit{on-shell}
amplitude in powers of the momentum, and fit to the scattering length
$\alpha$ and the effective range $r$,
\begin{eqnarray}
 \left.{\cal A}^{-1}\right|_{on-shell}&=&
  -\frac{M}{4\pi}
  \left[
   -\frac{1}{\alpha}+\half r p^2 +{\cal O}(p^4)-ip
  \right], \nonumber \\
 &&\mbox{\rm with}\ 
  p=\sqrt{Mp^0}=\left|\bfp_1\right|=\left|\bfp_2\right|
\end{eqnarray}
to determine the couping constants $C_0$ and $C_2$ as functions of the
cutoff $\Lambda$. It is however impossible to do so \textit{off-shell},
because $\mu$ is an independent variable. We have three equations for
the two coupling constants. In other words, the conventional method does
not renormalize the off-shell amplitude.

The notion of renormalization should be independent of whether we
renormalize the amplitude on-shell or off-shell. It is clear that the
troublesome terms contain the factor $\bfp_1^2+\mu^2$ or $\bfp_2^2+\mu^2$,
i.e., $\nabla^2+Mp^0$ in coordinate space. As we will show shortly, the
operator proportional to the ``equation of motion'' is necessary to
renormalize the theory off-shell.

In reality, the operators such as
\begin{equation}
 2B\left[\left(N^T N\right)^\dagger
  N^T\left(i\der_t+\frac{\nabla^2}{2M}\right)N
  + h.c. \right],
 \label{B-op}
\end{equation}
(where $B$ is a coupling constant) must be included in the EFT
Lagrangian, because they are local operators satisfying the symmetry of
the theory. But these operators are usually discarded by using the field
redefinition,
\begin{equation}
 N\rightarrow N-2BN^\dagger\left(N^T N\right).
  \label{redef}
\end{equation}
As emphasized in Sec.~\ref{sec:indep}, this field redefinition gives
rise to an extra factor coming from the measure, which should be treated
carefully in our case.

It would be possible to calculate the contributions from the measure
explicitly if we properly regularize the products of the operators in
the field redefinition and the measure itself. Here, we instead follow
an indirect way.

Because the transformation is local and does not contain derivatives,
the extra interaction coming from the measure must be represented as (an
infinite set of) local operators. Furthermore, such operators should
satisfy the symmetry of the theory. However, all the possible
interactions are already included in the EFT Lagrangian! The net effect
is therefore to change the coupling constants. We can determine the
changes of the coupling constants by demanding that the transformed
theory (without the B-interaction) have the same physical amplitude as
that of the original theory (with the B-interaction).

Let us consider the (``original'') theory with the B-interaction
(\ref{B-op}) included. The LS equation may be solved in the similar way,
with the vertex $V$ being replaced by
\begin{eqnarray}
 V'(p^0,\bfp_1,\bfp_2)&=&C'_0+4C'_2\left(\bfp_1^2+\bfp_2^2\right) 
  \nonumber \\
 &&{}-2B\left(p^0-\left(\bfp_1^2+\bfp_2^2\right)/2M\right)+\cdots.
  \nonumber \\
\end{eqnarray}
The amplitude
\begin{equation}
 {\cal A}'(p^0,\bfp_1,\bfp_2)=x'(p^0) 
  + y'(p^0)(\bfp_1^2+\bfp_2^2)+z'(p^0) \bfp_1^2\bfp_2^2,
\end{equation}
may be easily obtained by noting that $V'$ is obtained by substituting
$C_0 \rightarrow C_0'-2Bp^0$ and $4C_2\rightarrow 4C_2'+B/M$ into $V$,
resulting
\begin{eqnarray}
 x'\!&=&\!\frac{1}{D'}
  \left(
   \left(C_0'\!-\!2Bp^0\right)
   \!+\!\left(4C'_2+\frac{B}{M}\right)^2\!I_2
  \right), \\
 y'\!&=&\!\frac{1}{D'}\left(4C'_2+\frac{B}{M}\right)
  \left(1\!-\!\left(4C'_2+\frac{B}{M}\right)\!I_1\right), \\
 z'\!&=&\!\frac{1}{D'}\left(4C'_2+\frac{B}{M}\right)^2I_0,
\end{eqnarray}
with
\begin{eqnarray}
 D'&=&1-\left(C_0'-2Bp^0\right)I_0-2\left(4C_2'+B/M\right)I_1 \nonumber \\
 &&{}+\left(4C'_2+B/M\right)^2I_1^2-\left(4C'_2+B/M\right)^2I_0I_2.
  \nonumber \\
\end{eqnarray}
We now demand that the denominator $D'(C_0', C_2', B)$ be proportional
to $D(C_0, C_2)$ \textit{off-shell},
\begin{equation}
 D(C_0, C_2)=R(B) D'(C_0', C_2', B).
\end{equation}
We require the off-shell proportionality because the relation between
$(C_0, C_2)$ and $(C_0', C_2', B)$ should be independent of the on-shell
nature, and the normalization of the amplitude may be affected by field
transformations.

The requirement has the solution,
\begin{subequations}
 \label{RGeqs}
\begin{eqnarray}
 \!\!\!\!C_0\!\!&=&\!\! R\left(C_0'+\left(4C_2'+\frac{B}{M}\right)^2L_5\right) 
  \nonumber \\
 &&\!\!\!{}-\frac{L_5}{L_3^2}
  \left[
   1\!-\!\sqrt{R}\left(1\!-\!\left(4C'_2\!+\!\frac{B}{M}\right)\!L_3\!\right)
  \right]^2\!\!, \label{eqC0}\\
 \!\!\!\!C_2\!\!&=&\!\!\frac{1}{4L_3}\!
  \left[
    1-\sqrt{R}\left(1-\left(4C'_2+\frac{B}{M}\right)\!L_3\!\right)
  \right]\!, \label{eqC2}\\
 \!\!\!\!R\!\!&=&\!\!\left[1-2\frac{B}{M}L_3\right]^{-1}, \label{eqR}
\end{eqnarray}
\end{subequations}
where
\begin{equation}
 L_n=-\frac{1}{n}\frac{M\Lambda^n}{2\pi^2}.
\end{equation}
It is easy to show that the \textit{on-shell} amplitudes are actually
identical for both theories if the above condition is satisfied.

Some remarks are in order. Firstly, by setting $B=0$, the solution
reduces to the trivial one; $C_0=C_0'$, $C_2=C_2'$, and
$R=1$. Secondly, $C_0$ and $C_2$ are linear in $C_0'$ and
$C_2'$. Thirdly, $C_0$ contains $C_2'$. Finally, the coefficient of
$C_0'$ in $C_0$ and that of $C_2'$ in $C_2$ are not equal. The former
two are what one would expect, while the latter two are difficult to
understand. One would naively expect
\begin{equation}
 C_0=A(B)C_0'+\delta_{0}(B), \quad C_2=A(B)C_2'+\delta_{2}(B),
\end{equation}
because the changes of the coupling constants ($\delta_0$ and
$\delta_2$) come form the Jacobian for the field redefinition
(\ref{redef}), it should contain only the coupling constant $B$. (The
factor $A(B)$ comes from the ``wave function renormalization'' due to
the Jacobian contribution proportional to the kinetic term.) We suspect
that this dependence of the Jacobian on the coupling constants $C_0'$
and $C_2'$ comes from the definition of the composite operator
$N^\dagger\left(N^T N\right)$.

\section{RG equations}
\label{sec:RG}

Once we obtain the complete two particle scattering amplitude, it is
easy to derive the RG equations for $C'_0$, $C_2'$, and $B$, by
demanding (the inverse of) the amplitude is, when expanded in
$\bfp_1^2$, $\bfp_2^2$, and $\mu^2$, independent of the cutoff $\Lambda$.

Let us introduce the dimensionless coupling constants $\gamma_0$,
$\gamma_2$, and $\beta$ as,
\begin{equation}
 \gamma_0=\frac{M\Lambda}{2\pi^2}C_0',\quad
  \gamma_2=\frac{M\Lambda^3}{2\pi^2}4C_2', \quad
  \beta=\frac{\Lambda^3}{2\pi^2}B.
\end{equation}
We obtain the following RG equations,
\begin{subequations}
 \label{rgXYZ}
\begin{eqnarray}
 \Lambda\frac{dX}{d\Lambda}&=& -(1-X)(Y+3X^2)/X, \label{rgX}\\
 \Lambda\frac{dY}{d\Lambda}&=& Y(6X^3-5X^2+2XY-Y)/X^2, \label{rgY}\\
 \Lambda\frac{dZ}{d\Lambda}&=& Y^2/X^2-3Z+6XZ+2YZ/X,\label{rgZ}
\end{eqnarray}
\end{subequations}
where $ X=1+(\gamma_2+\beta)/3,\ Y = \gamma_0 - (\gamma_2+\beta)^2/5$,
and $Z = 2\gamma_2+(\gamma_2+\beta)^2/3$.  This set of equations has a
nontrivial fixed point $(X^*,Y^*,Z^*) = (1,-1,-1)$ corresponding to
\begin{equation}
 (\gamma_0^*, \gamma_2^*, \beta^*)=\left(-1,-\half,\half\right),
\end{equation}
beside the trivial one, $(X,Y,Z)=(1,0,0)$ corresponding to $(\gamma_0,
\gamma_2, \beta)=(0,0,0)$. At the nontrivial fixed point, the theory is
of course scale invariant\cite{Mehen:1999nd} and the scattering length
is infinite. In the real world, we are a bit away from the fixed point.

We may linearize the RG equations around the fixed point by substituting
\begin{equation}
 \gamma_0=\gamma_0^*+\epsilon_0,\quad
  \gamma_2=\gamma_2^*+\epsilon_2,\quad
  \beta=\beta^*+\epsilon_\beta,
\end{equation}
so that we have
\begin{subequations}
 \label{rglinear}
\begin{eqnarray}
 \Lambda\frac{d\epsilon_0}{d\Lambda}&=&
  -\epsilon_0-2\epsilon_2-2\epsilon_\beta,  \\
 \Lambda\frac{d\epsilon_2}{d\Lambda}&=&
  -2\epsilon_0-\frac{2}{3}\epsilon_2-\frac{5}{3}\epsilon_\beta, \\
 \Lambda\frac{d\epsilon_\beta}{d\Lambda}&=&
  2\epsilon_0+\frac{8}{3}\epsilon_2+\frac{11}{3}\epsilon_\beta.
\end{eqnarray}
\end{subequations}
These equations are easily solved. We have found the following
eigenvalues and the corresponding eigenvectors;
\begin{equation}
 2: \left(
     \begin{array}{c}
      2 \\
      1 \\
      -4
     \end{array}
    \right), \quad
 1: \left(
     \begin{array}{c}
      0\\
      -1 \\
      1
     \end{array}
    \right), \quad
 -1: \left(
     \begin{array}{c}
      1\\
      1 \\
      -1
     \end{array}
    \right).
\end{equation}
It is important to note that there is a negative eigenvalue $-1$; namely,
there is a relevant combination of the operators. 

By using them, we have 
\begin{subequations}
 \label{linearsol}
\begin{eqnarray}
 \epsilon_0&=&2a\left(\frac{\Lambda}{\Lambda_0}\right)^2
  +c\left(\frac{\Lambda_0}{\Lambda}\right), \\
 \epsilon_2&=&a\left(\frac{\Lambda}{\Lambda_0}\right)^2
  -b\left(\frac{\Lambda}{\Lambda_0}\right)
  +c\left(\frac{\Lambda_0}{\Lambda}\right), \\
 \epsilon_\beta&=&-4a\left(\frac{\Lambda}{\Lambda_0}\right)^2
  +b\left(\frac{\Lambda}{\Lambda_0}\right)
  -c\left(\frac{\Lambda_0}{\Lambda}\right),
\end{eqnarray}
\end{subequations}
where $a$, $b$, and $c$ are infinitesimal dimensionless constants and
$\Lambda_0$ is a parameter of mass dimension. Inserting these solutions,
we obtain the (renormalized) off-shell amplitude near the nontrivial
fixed point,
\begin{equation}
 \left.{{\cal A}'}^{-1}(p^0, \bfp_1,\bfp_2)\right|_*
  =-\frac{M\Lambda_0}{2\pi^2}c-\frac{Mb}{\pi^2\Lambda_0}\mu^2
  -\frac{M\mu}{4\pi}+\cdots,
\end{equation}
where ellipsis denotes higher orders in $a$, $b$, and $c$. It is
important to note that the pole is independent of the relative momenta
$\bfp_1$ and $\bfp_2$ of the center-of-mass system, as it should
be. This point is not clear from the (renormalized) on-shell amplitude.

Finally we describe the method of obtaining the RG equations for the
reduced set of coupling constants. Start with the set of the coupling
constants $(C_0, C_2)\equiv(C_0', C_2', B=0)$, and infinitesimally lower
the cutoff $\Lambda$ by $\delta \Lambda$. It transforms $(C_0', C_2',
B=0)$ to $(C_0'+\delta C_0', C_2'+\delta C_2', \delta B)$ according to
the RG equations (\ref{rgXYZ}). Now we invoke the equivalence relation
(\ref{RGeqs}) to eliminate $\delta B$;
\begin{equation}
 (C_0'+\delta C_0', C_2'+\delta C_2', \delta B)
  \rightarrow (C_0+\delta C_0, C_2+\delta C_2).
\end{equation}
The resulting RG equations are identical to those obtained from the
$\Lambda$-independence of the \textit{on-shell} amplitude, as expected.

\section{Conclusion}
\label{sec:conclusion} 

In this paper, we showed that the elimination of redundant operators
generically has the Jacobian contribution and that it plays an important
role in the Wilsonian RG analysis. The ``pionless'' EFT was considered
as a concrete example. It was shown that the \textit{off-shell}
amplitudes cannot be renormalized without the redundant operator (the
``B-interaction''). Off-shell renormalization is important, particularly
in analyzing the three-body systems.

We established the equivalence relation between the theories with and
without the redundant operator. The Jacobian contribution was shown to
have very peculiar features.

We performed a Wilsonian RG analysis based on the two particle scattering
amplitude. It is important to note that a Wilsonian RG analysis is
impossible without the redundant operators, because the RG
transformations generate all possible operators which satisfy the
symmetry of the theory, including the redundant operators.

We derive the equivalence relation by directly comparing the
(nonperturbative) amplitudes, but in many cases it is impossible to
obtain them. A direct method of calculating the Jacobian factor is desired.

The physical significance of the Wilsonian RG analysis of Nuclear EFT
will be reported elsewhere\cite{HK}. Here we just mention that the RG
flow determines the \textit{phase structure} of the nuclear force and
the inverse of the scattering length may be identified with the
\textit{order parameter}; it characterizes the relevant direction near
the nontrivial fixed point.

\begin{acknowledgments}
 The authors would like to thank M.~C.~Birse for e-mail correspondences.
 One of the authors (K.H.) is partially supported by Grant-in-Aid for
 Scientific Research on Priority Area, Number of Area 763, ``Dynamics of
 Strings and Fields,'' from the Ministry of Education, Culture, Sports,
 Science and Technology, Japan. Another (K.I.) is partially supported by
 Grant-in-Aid for Scientific Research on Priority Area, Number of Area
 441, ``Progress in Elementary Particle Physics of the 21th Century
 though Discoveries of Higgs Boson and Supersymmetry'' from the Ministry
 of Education, Culture, Sports, Science and Technology, Japan.
\end{acknowledgments}

\bibliography{NEFT,NPRG}

\end{document}